\documentclass{INTERSPEECH2023}

\interspeechcameraready

\usepackage{multirow}

\title{Language-specific Acoustic Boundary Learning for Mandarin-English Code-switching Speech Recognition}
\name{Zhiyun Fan$^1$, Linhao Dong$^1$, Chen Shen$^1$, Zhenlin Liang$^1$, Jun Zhang$^1$, Lu Lu$^1$, Zejun Ma$^1$}
\address{
  $^1$Bytedance Research }
\email{\{fanzhiyun.20203,donglinhao,shenchen.0622,liangzhenlin.lzl\}@bytedance.com}

\begin{document}

\maketitle
 
\begin{abstract}

Code-switching speech recognition (CSSR) transcribes speech that switches between multiple languages or dialects within a single sentence. The main challenge in this task is that different languages often have similar pronunciations, making it difficult for models to distinguish between them. In this paper, we propose a method for solving the CSSR task from the perspective of language-specific acoustic boundary learning. We introduce language-specific weight estimators (LSWE) to model acoustic boundary learning in different languages separately. Additionally, a non-autoregressive (NAR) decoder and a language change detection (LCD) module are employed to assist in training. Evaluated on the SEAME corpus, our method achieves a state-of-the-art mixed error rate (MER) of 16.29\% and 22.81\% on the $\text{test}_{\text{man}}$ and $\text{test}_{\text{sge}}$ sets. We also demonstrate the effectiveness of our method on a 9000-hour in-house meeting code-switching dataset, where our method achieves a relatively 7.9\% MER reduction.
\end{abstract}

\section{Introduction}
The primary objective of code-switching (CS) is to facilitate effective communication across diverse linguistic or technical domains. CS entails the practice of alternating between two or more languages in a single sentence. However, incorporating words or phrases from multiple languages can result in transcription errors and confusion, which makes code-switching speech recognition (CSSR) a more challenging task \cite{sitaram2019survey}. 

The CSSR task has been studied for decades. In the early days, most works are conducted on the hybrid framework \cite{adel2014features,adel2015syntactic,adel2013combination,vu2012first,yilmaz2018acoustic,guo2018study}. However, as the end-to-end (E2E) model become increasingly popular, researchers begin pursuing an end-to-end strategy to resolve the CSSR task. The attention-based E2E models are first applied to the CSSR task, and to improve speech recognition performance, language identification (LID) is used as an auxiliary task \cite{zeng2018end,luo2018towards}. Additionally, the language-aware encoder (LAE) structures and language-aware training (LAT) are applied to connectionist temporal classification (CTC) and neural transducer systems to disentangle language-specific information and generate frame-level language-aware representations during encoding \cite{tian2022lae,song2022monolingual,song2022language,dalmia2021transformer,yan2022towards,zhang2021rnn,li2019towards}. For the decoder, language-related attention mechanisms \cite{zhang2022reducing}, non-autoregressive structures \cite{peng2022minimum,chuang2021non} and internal language model estimation (ILME) \cite{peng2022internal} based language models are all used to alleviate the confusion brought by the code-switching of different languages. To model both the monolingual and the cross-lingual sequential dependency, Lee et al. propose a bilingual attention language model (BALM) that simultaneously performs language modeling objectives with a quasi-translation objective \cite{lee2020modeling}. Multi-encoder-decoder (MED) explores the language-related structure on both the encoder and decoder \cite{zhou2020multi}. 

In general, the E2E ASR model consists of the encoder, decoder, and alignment mechanism. Most of the existing E2E CSSR models only focus on optimizing the encoder and decoder structure, while few works explore whether the alignment mechanism needs to be language-specific. On the other hand, most previous works use a mixture of Mandarin characters and English subwords as modeling units \cite{zeng2018end,luo2018towards,tian2022lae,song2022monolingual,song2022language,dalmia2021transformer,yan2022towards,zhang2022reducing,peng2022minimum,zhou2020multi}. Mandarin character typically represents a single syllable in Mandarin Chinese \cite{yang2002acoustic}, and their acoustic boundary is clear \cite{dong2020cif}. However, subwords are obtained  without referring to any acoustic knowledge \cite{sennrich2015neural}, and their acoustic boundary may be blurred \cite{dong2020cif}. To obtain good acoustic boundaries (alignment) for both Mandarin and English in the CSSR system, it is intuitive to conduct language-specific boundary learning.

%Yan et al. \cite{yan2022towards} split the cs task into monolingual stage and CS point detection. 

% The indiscriminate transliterator used in \cite{yan2022towards} inspires us that these confusion in CSSR task clould be alleviated by boundary optimization. We borrow an example from \cite{yan2022towards} to explain it in Fig. \ref{fig:example}. For a speech of 'accounting', the right English transcripts respond to two speech segments for subwords, and the wrong Mandarin transcripts respond to three segments for the character. After further literature research, we found the generation of bpe does not refer to any pronunciation information, and characters used in Mandarin are of clear acoustic boundary. So, language-specific boundary learning is necessary when the output units are combined with Mandarin characters and English BPE. We found some similar bad cases in our baseline model, and our proposed method improve these cases as we expected. In the experiments, we give out a presentation of it. 

% \begin{figure}[t]
%   \centering
%   \includegraphics[width=\linewidth]{example.pdf}
%   \caption{Schematic diagram of speech production.}
%   \label{fig:example}
% \end{figure}
 
In this paper, we employ the CIF-based model \cite{dong2020cif} as our backbone, which utilizes a weight estimator to predict acoustic boundaries for aligning the encoder and decoder. For the CSSR task, we introduce language-specific weight estimators (LSWE) to enable language-specific boundary learning for different languages. Additionally, a non-autoregressive decoder is utilized to assist in learning the language-specific boundaries. To further enhance the model's ability to detect intra-sentence language changes, we design a language change detection (LCD) module. It's worth noting that the non-autoregressive decoder and LCD module are eliminated in the inference stage, which means our model has almost no increase in the number of parameters compared to the vanilla CIF-based model. We evaluate the proposed method on two datasets: SEAME, a public conversational Mandarin-English corpus, and an in-house meeting code-switching dataset. On the SEAME benchmark, our method outperforms strong baselines and achieves a new state-of-the-art performance, obtaining an MER of $16.29\%$ and $22.81\%$ on the $\text{test}_{\text{man}}$ and $\text{test}_{\text{sge}}$, respectively. On the in-house meeting CS dataset with 9000 hours of speech, our method shows a relative MER reduction of $7.9\%$, further validating its effectiveness in real-world scenarios. As far as we know, this work is the first to consider language-specific boundary learning in the CSSR task.

% select cif model as the base framework and conduct an exploration of the language-specific boundary learning method. Firstly, cif is a E2E ASR framework, which consists of the encoder, decoder, weight estimator and cif module. The last two parts complete the alignment of the encoder to the decoder. The boundary is decided by the weights estimator outputs. So, for conducting language-specific boundary learning, we design two weight estimators for English and Mandarin, respectively. The mixed weight is obtained by adding the weights of Mandarin and English. In order to enable the two weight estimators to learn the boundary patterns of their respective languages, we designed a non-autoregressive decoder on the backend to calculate the monolingual cross-entropy loss. We additionally design a language classification module to enhance the ability of our model to make intral-sentence language predictions. 

\begin{figure*}[t]
  \centering
  \includegraphics[width=0.79\linewidth]{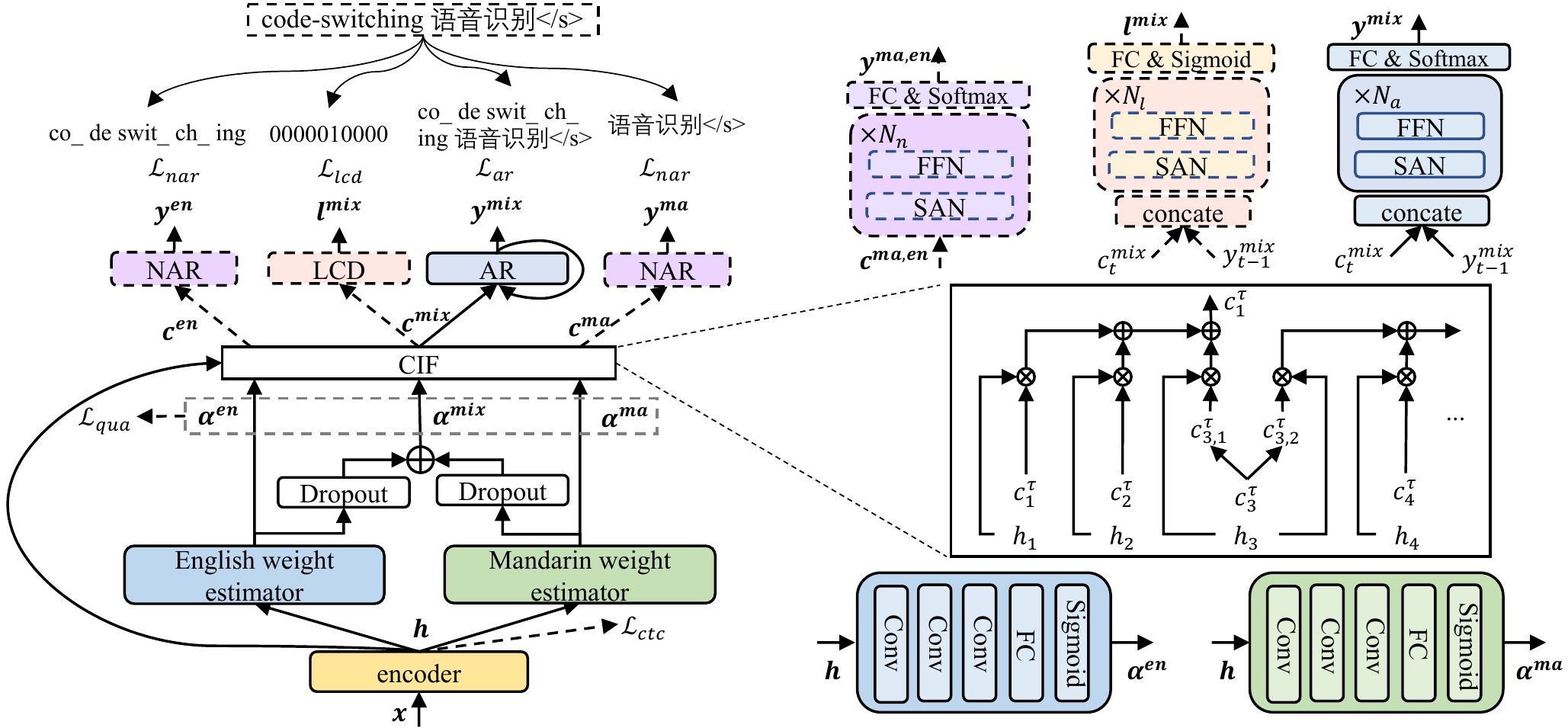}
\setlength{\abovecaptionskip}{0.1cm}
  \caption{Schematic diagram of the proposed method. The left part illustrates the overall structure, and the right part provides details on utilized modules. Modules with the same color correspond to each other in the left and right parts. $\tau$ in the CIF mechanism can be ma, en, or mix. In the upper left corner, there is an example of generating monolingual labels from the target sequences. The $\langle /s \rangle$ is assigned to the language that corresponds to the last token. The modules with dotted borders are eliminated in the inference stage.}
  \vspace{-0.5cm}
  \label{fig:framework}
\end{figure*}
\vspace{-0.12cm}
\section{Continuous and integrate-and-fire}
Continuous integrate-and-fire (CIF) is an alignment mechanism employed in the E2E ASR model \cite{dong2020cif}. Typically, it comes equipped with a weight estimator that calculates the acoustic information weight in each frame. The schematic diagram of CIF calculation is presented in the middle right part of Figure \ref{fig:framework}. CIF sequentially accumulates the acoustic information weight $\bm{c}$, and once the accumulated weights reach a threshold $\beta$, the corresponding frames are located as acoustic boundaries. All frame-level encoded outputs $\bm{h}$ between every two adjacent boundaries are then weighted and summed using the information weights to obtain a token-level acoustic embedding $\bm{c}$. The information weights corresponding to boundary frames are divided into two parts to ensure that the sum of information weight between two adjacent boundaries equals 1.

In summary, the CIF mechanism can convert the frame-level acoustic representation to the token level using the weight information predicted by the weight estimator. However, the CIF mechanism involves mathematical calculations and does not have trainable parameters. Therefore, the weight estimator is the only trainable component that ultimately determines the location of the acoustic boundaries. Thus, all our language-specific boundary learning is focused on the weight estimator.
\vspace{-0.12cm}
\section{Proposed method}
\vspace{-0.12cm}
\subsection{Overview}
This article builds upon the CIF model \cite{dong2020cif} and proposes improvements to enable language-specific boundary learning for the CSSR task. The details of the model architecture are presented in Figure \ref{fig:framework}. The model consists of six components: an encoder, language-specific weight estimators (LSWE), a CIF calculator, an autoregressive (AR) decoder, a non-autoregressive (NAR) decoder, and a language change detection (LCD) module. The input feature sequence $\bm{x}$ is transformed into an encoded output $\bm{h}$ by the encoder, which comprises conformer layers \cite{gulati2020conformer}. The encoded output $\bm{h}$ is passed through two language-specific weight estimators, which predict the acoustic boundary and information contained in each frame. The predictions for Mandarin and English are denoted as $\bm{\alpha^{ma}}$ and $\bm{\alpha^{en}}$, respectively. The mixture weight $\bm{\alpha}^{mix}$ is the frame-level addition of $\bm{\alpha^{ma}}$ and $\bm{\alpha^{en}}$. The CIF calculator utilizes the language-specific information weight $\bm{\alpha^{ma}}$, $\bm{\alpha^{en}}$, and $\bm{\alpha^{mix}}$ to convert the frame-level acoustic representation $\bm{h}$ into token-level acoustic embeddings, namely $\bm{c^{ma}}$, $\bm{c^{en}}$, and $\bm{c^{mix}}$ for Mandarin, English, and the mixture, respectively. The AR decoder takes in acoustic embedding $\bm{c^{mix}}$ to generate target prediction $\bm{y^{mix}}$. The NAR decoder takes in monolingual acoustic embeddings $\bm{c^{ma}}$ and $\bm{c^{en}}$ to predict the corresponding monolingual token sequence $\bm{y^{ma}}$ and $\bm{y^{en}}$. Lastly, the LCD module takes in $\bm{c^{mix}}$ and the previously predicted token from the AR decoder to predict language change probability $\bm{l^{mix}}$.

During inference, the NAR decoder and LCD module are not utilized, resulting in almost identical parameter numbers between our method and the baseline model. Moreover, the calculation of monolingual acoustic embeddings $\bm{c^{ma}}$ and $\bm{c^{en}}$ is omitted. The primary role of the NAR decoder and LCD module is to assist in the learning of $\bm{\alpha^{ma}}$ and $\bm{\alpha^{en}}$, which are strongly correlated with the acoustic boundary. 
 %The two weight estimators are of the same structure, which stacks three convolution layers and a full-connected layer with a sigmoid activation function. 
 
%The autoregressive decoder stacks self-attention (SAN), feed-forward networks (FFN) and a softmax output layer. It responds for predicting the target sequence. The non-autoregressive decoder is of the same structure as the autoregressive decoder except that the input doesn't concate the token prediction from the last decoding step. It uses the monolingual CE loss to assist the boundary learning of corresponding languages. The language change detection module stacks SAN and FFN. It uses the current acoustic embedding and the token prediction from the last decoding step to predict whether current there is a language change happening. It aims at enhancing the ability of LWE to detect language change. 
\vspace{-0.12cm}
\subsection{Language-specific weight estimator}
In the vanilla CIF-based ASR model \cite{dong2020cif}, the information weight predicted by the weight estimator can decide the acoustic boundary through the CIF calculation. To achieve language-specific boundary learning, we design two weight estimators for Mandarin and English, respectively. The $\bm{\alpha^{ma}}$ and $\bm{\alpha^{en}}$ represent the Mandarin and English information contained in each frame and determine the location of the acoustic boundary. The mixed information weight is fused from the two information weight as follow. 
\begin{equation}
\setlength{\abovedisplayskip}{3pt} 
\setlength{\belowdisplayskip}{3pt}
  \bm{\alpha^{mix}} = \text{Dropout}(\bm{\alpha^{ma}}, p) + \text{Dropout}(\bm{\alpha^{en}}, p)
\end{equation}
where $p$ represents the dropout rate \cite{srivastava2014dropout}. To prevent the model from relying too heavily on either language and encourages it to learn from both, we apply dropout to the information weights of both languages before adding them. Additionally, we modify the scaling operation used during training in the vanilla CIF-based model to be language-specific.
\begin{equation}
\setlength{\abovedisplayskip}{3pt} 
\setlength{\belowdisplayskip}{3pt}
  \bm{\alpha^{\tau'}} = \bm{\alpha^{\tau}} *U^{\tau} /  \sum_{t=1}^{T'}  \bm{\alpha^{\tau}} 
\end{equation}
where the $\tau$ can be ma, en or mix. 

The quantity loss function is utilized to train the language-specific weight estimator, enabling it to accurately predict the number of tokens in a sentence for the corresponding languages.
\begin{equation}
\setlength{\abovedisplayskip}{3pt} 
\setlength{\belowdisplayskip}{3pt}
\begin{aligned}
  \mathcal{L}_{qua}= & |U^{mix} - \sum_{t=1}^{T'} \bm{\alpha^{mix}}| \\
  & + \frac{1}{2}(|U^{ma} - \sum_{t=1}^{T'} \bm{\alpha^{ma}}| + |U^{en} - \sum_{t=1}^{T'} \bm{\alpha^{en}}|)
\end{aligned}
\end{equation}
where the $U^{ma}$, $U^{en}$, and $U^{mix}$ represent the token numbers of Mandarin, English, and mixture, respectively. $T'$ represents the length of frame-level representation $\bm h$.
\vspace{-0.12cm}
\subsection{Auxiliary modules}
%We design a non-autoregressive decoder and a language change detection module to assist in the learning of the language-specific weight estimator. 
The non-autoregressive decoder solely processes monolingual acoustic embeddings and utilizes a monolingual cross-entropy loss to assist in the learning of LSWE. The degradation of $\mathcal{L}_{nar}$ guides the LSWE to predict better acoustic boundaries for monolingual acoustic embeddings $\bm{c^{ma}}$ and $\bm{c^{en}}$.
\begin{equation}
\setlength{\abovedisplayskip}{3pt} 
\setlength{\belowdisplayskip}{3pt}
    \mathcal{L}_{nar} = -\sum_{t=1}^{U^{ma}}\text{log}p(y^{ma}_t)|y^{ma}_{< t},c^{ma}_{\leq t})-\sum_{t=1}^{U^{en}}\text{log}p(y^{en}_t)|y^{en}_{< t},c^{en}_{\leq t})
\end{equation}

The monolingual labels that supervise the NAR decoder are generated by splitting the target sequence on the fly. An example of the monolingual label generation process is depicted in the upper left corner of Figure \ref{fig:framework}. It should be noted that in contrast to previous works \cite{tian2022lae,song2022monolingual}, our monolingual labels do not include a placeholder for the other language.

The LCD module is employed to improve the model's capacity to detect language change within a sentence. This is achieved by incorporating a token-level binary cross-entropy loss.
\begin{equation}
\setlength{\abovedisplayskip}{3pt} 
\setlength{\belowdisplayskip}{3pt}
\begin{aligned}
    \mathcal{L}_{lcd}=&-\sum_{t=1}^{U^{mix}}\left\{\hat{l}^{mix}_t\text{log}p(l^{mix}_t|y^{mix}_{< t},c^{mix}_{\leq t}) \right. \\
    & \left. +(1-\hat{l}_t^{mix})\text{log}\left[1-p(l^{mix}_t|y^{mix}_{< t},c^{mix}_{\leq t})\right]\right\}
\end{aligned}
\end{equation}
where $\bm{\hat{l}^{mix}}$ represents the ground truth label for language change detection.
\vspace{-0.12cm}
\subsection{Loss function}
We employ a joint training strategy to optimize all parameters, where the overall objective function is a weighted sum of all the losses mentioned above.
\begin{equation}
\setlength{\abovedisplayskip}{3pt} 
\setlength{\belowdisplayskip}{3pt}
    \mathcal{L}=\mathcal{L}_{ar} + \lambda_{ctc}\mathcal{L}_{ctc} + \lambda_{qua}\mathcal{L}_{qua} + \lambda_{nar}\mathcal{L}_{nar} + \lambda_{lcd}\mathcal{L}_{lcd}
\end{equation}
where $\lambda_{ctc}$, $\lambda_{qua}$, $\lambda_{nar}$ and $\lambda_{lcd}$ are tunable hyper-parameters.
\vspace{-0.12cm}
\section{Experiments}

\begin{figure*}[t]
  \centering
  \includegraphics[width=\linewidth]{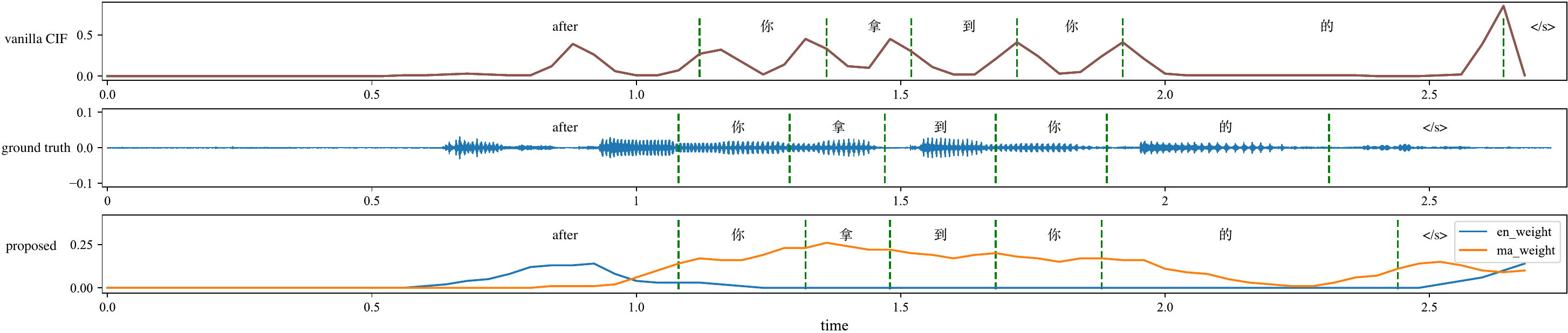}
  \setlength{\abovecaptionskip}{-0.1cm}
  \caption{The upper part of the figure shows the weight estimator output in the vanilla CIF model, while the lower part shows the output of our two language-specific weight estimators. The green dotted lines mark the boundaries. The figure is for the utterance index $\text{46NC41MBP\_0101\_3035860\_3038590}$ in the $\text{test}_{\text{sge}}$ test set.}
  \vspace{-0.5cm}
  \label{fig:weight}
\end{figure*}
\vspace{-0.12cm}
\subsection{Datasets}
We perform experiments on two datasets: SEAME \cite{lyu2010seame} and an in-house meeting code-switching dataset. SEAME comprises approximately 105 hours of spontaneous Mandarin-English intra-sentential code-switching speech from Singapore and Malaysia. Two evaluation subsets, biased towards Mandarin and Southeast Asian English, are denoted as $\text{test}_{\text{man}}$ and $\text{test}_{\text{sge}}$, respectively. Due to space constraints, we refer readers to \cite{peng2022internal} for further details on SEAME. The in-house meeting code-switching dataset is collected from the recording of real meetings and manually transcribed. It comprises 3000 hours of monolingual English training data, 3000 hours of monolingual Mandarin training data, and 3000 hours of Mandarin-English code-switching training data. Two monolingual and one code-switching evaluation sets are utilized, and they are denoted as $\text{test}_{\text{ma}}$, $\text{test}_\text{n}$, and $\text{test}_\text{cs}$, respectively. %The details are shown in Table \ref{tab:data}. 
They comprise 8.9, 10.78, and 10.93 hours of speech, respectively.

% \begin{table}[bt!]
%   \caption{Details of the in-house meeting code-switching dataset. "Man/Eng" represents the ratio of Mandarin characters to English subwords for specific data sets.}
%   \label{tab:data}
%   \centering
%   \begin{tabular}{lcccc}
%     \toprule
%     \textbf{Details} &  \textbf{train} & $\textbf{test}_{\textbf{ma}}$&  $\textbf{test}_{\textbf{en}}$ & $\textbf{test}_{\textbf{cs}}$ \\
%          \midrule
%         \# Utterances & 6488569  & 6217 & 5842 & 4072 \\
%         Duration(Hrs) & 9000 & 8.95 & 10.78 & 10.93  \\
%         Man/Eng & 0.77/0.23 & 1.0/0.0 & 0.0/1.0 & 0.93/0.07 \\
%     \bottomrule
%   \end{tabular}
% \end{table}
\vspace{-0.12cm}
\subsection{Experimental settings}
We use 80-dimensional log-Mel filter-bank features, computed with a 25 ms window and shifted every 10 ms, for all experiments. A convolutional layer with 64 filters and 1/2 temporal downsampling is used as the front end of the encoder. The encoder comprises 15 conformer \cite{gulati2020conformer} layers with 4 attention heads, 256 attention dimensions, and 1024 feed-forward network (FFN) dimensions. There is a max-pooling layer for 1/2 temporal downsampling after the eighth layers. The Mandarin and English weight estimators have identical structures, consisting of three 1-dimensional convolutional layers and an FC layer. The kernel size of the convolutional layer is set to (3,1,3), and the filter number is 256. The FC layer has one output unit with sigmoid activation. The autoregressive decoder, non-autoregressive decoder, and LCD module have the same structure, consisting of 3 transformer \cite{vaswani2017attention} layers with 4 attention heads, 256 attention dimensions, and 1024 FFN dimensions. The only difference is in the input projection layers. The threshold $\beta$ used in the CIF mechanism is set to 1. The modeling units comprise 1686 English BPE subwords \cite{sennrich2015neural} and 2641 Mandarin characters. All the dropout rate is set to 0.1. The hyper-parameter $\lambda_{ctc}$, $\lambda_{qua}$, $\lambda_{nar}$, and $\lambda_{lcd}$ are set to 0.5, 0.01, 0.2, 0.1, respectively.

In the training stage, we employ the Adam optimizer \cite{kingma2014adam}. The learning rate warms up for the first 1k iterations to a peak of $10^{-3}$ and holds on for the next 40k iterations, and then linearly decays to $10^{-4}$ for the last 30k iterations. The last 10 checkpoints are averaged to obtain the final model for evaluation. For decoding, we employ beam-search \cite{graves2012sequence} with a beam size of 10. 
\vspace{-0.12cm}
\subsection{Metrics}
We use mix error rate (MER) as the evaluation metric for speech recognition. For monolingual cases, we use character error rate (CER) for Mandarin and word error rate (WER) for English. To evaluate the quality of the token acoustic boundary, we use the f1-score (F1).
% \begin{equation}
%     \text{BCR}=\frac{\#\text{sentences with correct boundary number prediction}}{\#\text{all sentences}}
% \end{equation}
\begin{equation}
\setlength{\abovedisplayskip}{3pt} 
\setlength{\belowdisplayskip}{3pt}
    \text{F1}=\frac{2*\text{Precision}*\text{Recall}}{\text{Precision}+\text{Recall}}
\end{equation}
\begin{equation}
\setlength{\abovedisplayskip}{3pt} 
\setlength{\belowdisplayskip}{3pt}
    \text{Precision}=\frac{\#\text{hit reference boundary}}{\#\text{reference boundary}}
\end{equation}
\begin{equation}
\setlength{\abovedisplayskip}{3pt} 
\setlength{\belowdisplayskip}{3pt}
    \text{Recall}=\frac{\#\text{hit hypothesis boundary}}{\#\text{hypothesis boundary}}
\end{equation}
where the "hit hypothesis boundary" means that at least one reference boundary falls within a tolerance window before and after the hypothesis boundary. Similarly, "Hit reference boundary" means that at least one predicted boundary falls within the same tolerance window before and after the reference boundary. For all our evaluations, the tolerance is set to 50 ms.
\vspace{-0.12cm}
\subsection{Results}
Table \ref{tab:base} presents a comparison between our model and a few strong baseline models on SEAME $\text{test}_{\text{man}}$/$\text{test}_{\text{sge}}$. Among these baselines, the result of CIF is our implementation of the vanilla CIF-based model \cite{dong2020cif}, and the conformer-AED \cite{peng2022internal} is the state-of-the-art model before our proposed method. In terms of the F1, our proposed method improves the acoustic boundary prediction compared to the CIF baseline. We attribute these improvements to the language-specific weight estimators. Figure \ref{fig:weight} visualizes the acoustic boundaries and information weight predicted by our Mandarin and English weight estimators. The information weight predicted by our LSWE has obvious language discrimination, and our proposed method gives better boundary predictions. The results of MER show that our method has achieved a significant improvement on both the two test sets of SEAME corpus compared with the strong baselines, and set a new state-of-the-art performance.  
% 增加可视化图的说明
\begin{table}[bt!]
  \caption{The overall performance (MER \%) and boundary metric (PR \%) of our proposed method on the SEAME corpus. Results from previous papers and our implementation of the vanilla CIF model are shown for comparison.}
  \label{tab:base}
  \centering
  \tabcolsep=0.15cm
  \begin{tabular}{lcccccc}
    \toprule
\multirow{2}{*}{\textbf{Method}} & \multicolumn{2}{c}{$\textbf{test}_{\textbf{man}}$} & \multicolumn{2}{c}{$\textbf{test}_{\textbf{sge}}$}   \\
\cmidrule(lr){2-3}\cmidrule(lr){4-5}
     & F1   $\uparrow$ & MER $\downarrow$ &  F1  $\uparrow$ & MER $\downarrow$  \\
         \midrule
         Baseline \\ 
         LF-MMI \cite{zeng2018end} & - & 22.10 &  - & 30.90  \\
         LSTM-AED \cite{zeng2018end}  & - & 25.60 &   - & 37.00  \\
         Transducer \cite{dalmia2021transformer}  & - & 19.20 &  -  & 26.90  \\
         Transformer-AED \cite{peng2022minimum}  & - & 17.89   & -  & 25.10  \\
        Conformer-AED \cite{peng2022internal}  & - & 16.40  & - &  23.20  \\
        Vanilla CIF \cite{dong2020cif} & 40.09 & 16.96 &   47.23 & 23.73     \\ 
          \midrule
    Our proposed     & \textbf{50.64} & \textbf{16.29} & \textbf{56.56}  & \textbf{22.81}      \\
    \bottomrule
  \end{tabular}
\end{table}

Our proposed method includes three modifications compared to the vanilla CIF-based model: language-specific weight estimators (LSWE), a NAR decoder trained with the monolingual label, and a LCD module. To evaluate the importance of these modifications, we conduct an ablation study on the SEAME corpus and the in-house meeting code-switching dataset. Table \ref{tab:abla} shows the impact of each change on our method. The results indicate that LSWE is the most important modification, as removing it causes the largest performance degradation on both datasets. Secondly, ablating the NAR decoder also results in a large performance loss, while the LCD module brings some benefits to our method but is not as crucial as other modifications. In addition, our results on the in-house meeting code-switching datasets show that our method achieves a relative MER reduction of 2.97\%, 6.4\%, and 7.9\% on the Mandarin, English, and code-switching test sets, respectively, compared with the vanilla CIF model (the result in the last row of Table \ref{tab:abla}). This indicates that our method improves the performance of the model in the CSSR task, particularly for code-switching speech.
 
\begin{table}[bt!]
  \caption{MER(\%) of ablation study on the SEAME corpus and the meeting code-switching dataset. Starting from our proposed full model, we remove its blocks and move towards a vanilla CIF-based model: (1) removing the LCD module; (2) removing the NAR decoder; (3) replacing the two LSWE with a shared weight estimator.}
  \label{tab:abla}
  \centering
    \tabcolsep=0.16cm
  \begin{tabular}{lcc|ccc}
    \toprule
    \textbf{Model} &   $\textbf{test}_{\textbf{man}}$ & $\textbf{test}_{\textbf{sge}}$   & $\textbf{test}_{\textbf{ma}}$&  $\textbf{test}_{\textbf{en}}$ & $\textbf{test}_{\textbf{cs}}$  \\
         \midrule
        Full model & 16.29 & 22.81 & 10.43 & 18.86  & 6.79 \\
         \ \  w/o LCD & 16.68 & 22.89 &  10.36 & 18.98 & 6.74  \\
      \ \  \ \ w/o NAR & 16.94 &  23.33 & 10.43 & 19.23 & 6.89  \\
     \  \ \  \ \ \  w/o LSWE & 16.96 & 23.73 & 10.75 & 20.15 & 7.37 \\
    \bottomrule
  \end{tabular}
\end{table}

\begin{table}[bt!]
  \caption{The results of our method when the NAR module takes in different token-level acoustic embedding. Mandarin and English use WER (\%) and CER (\%), respectively, while MER (\%) is used for mix.}
  \label{tab:inpnar}
  \centering
  \tabcolsep=0.12cm
  \begin{tabular}{lcccccccc}
    \toprule
\multirow{2}{*}{\textbf{Model}} & \multicolumn{3}{c}{$\textbf{test}_{\textbf{man}}$} & \multicolumn{3}{c}{$\textbf{test}_{\textbf{sge}}$}   \\
\cmidrule(lr){2-4}\cmidrule(lr){5-7}
     & MA & EN  &  ALL  & MA & EN  &  ALL   \\
         \midrule
    $\varnothing$ $\rightarrow$ NAR &14.19&24.57&16.85&18.06&26.21&23.20     \\
    Ma $\rightarrow$ NAR &13.71&24.58&16.50&17.65&26.25&23.07     \\
    En $\rightarrow$ NAR & 14.04&23.91 &16.58&18.08 &25.71&22.89     \\
    Ma,En $\rightarrow$ NAR & 13.60 & 24.07&16.29&17.78&25.75&22.81    \\
    \bottomrule
  \end{tabular}
\end{table}

In the full model of our method, the NAR decoder takes in both Mandarin and English token-level monolingual acoustic embedding to assist in learning the two language-specific weight estimators. Table \ref{tab:inpnar} shows the results of ablating the NAR decoder inputs. When we feed no input to the NAR decoder, the model obtains unsatisfactory results for both Mandarin and English in the two evaluation sets. However, when the NAR decoder takes in Mandarin or English token-level acoustic embedding, the recognition performance of our method for the corresponding language improves. The best setting is to feed both Mandarin and English token-level acoustic embeddings to the NAR decoder. The results in Table \ref{tab:inpnar} demonstrate that the NAR decoder can use the monolingual token-level acoustic embeddings generated by the LSWE to independently or simultaneously optimize the recognition performance of Mandarin and English. 
\vspace{-0.12cm}
\section{Conclusion}
This paper employs the CIF model to address the CSSR task from the perspective of language-specific boundary learning. We introduce two language-specific weight estimators designed to handle boundary learning for Mandarin and English separately. Additionally, we introduce a NAR decoder and an LCD module to facilitate the training of the language-specific weight estimators. Our experimental results on the SEAME corpus and an in-house meeting code-switching dataset demonstrate that the proposed method improves the model's boundary prediction and speech recognition performance for both Mandarin and English.

\bibliographystyle{IEEEtran}
\bibliography{mybib}

\begin{thebibliography}{10}
\providecommand{\url}[1]{#1}
\csname url@samestyle\endcsname
\providecommand{\newblock}{\relax}
\providecommand{\bibinfo}[2]{#2}
\providecommand{\BIBentrySTDinterwordspacing}{\spaceskip=0pt\relax}
\providecommand{\BIBentryALTinterwordstretchfactor}{4}
\providecommand{\BIBentryALTinterwordspacing}{\spaceskip=\fontdimen2\font plus
\BIBentryALTinterwordstretchfactor\fontdimen3\font minus
  \fontdimen4\font\relax}
\providecommand{\BIBforeignlanguage}[2]{{%
\expandafter\ifx\csname l@#1\endcsname\relax
\typeout{** WARNING: IEEEtran.bst: No hyphenation pattern has been}%
\typeout{** loaded for the language `#1'. Using the pattern for}%
\typeout{** the default language instead.}%
\else
\language=\csname l@#1\endcsname
\fi
#2}}
\providecommand{\BIBdecl}{\relax}
\BIBdecl

\bibitem{sitaram2019survey}
S.~Sitaram, K.~R. Chandu, S.~K. Rallabandi, and A.~W. Black, ``A survey of
  code-switched speech and language processing,'' \emph{arXiv preprint
  arXiv:1904.00784}, 2019.

\bibitem{adel2014features}
H.~Adel, K.~Kirchhoff, D.~Telaar, N.~T. Vu, T.~Schlippe, and T.~Schultz,
  ``Features for factored language models for code-switching speech,'' in
  \emph{Proc. Spoken Language Technologies for Under-Resourced Languages
  (SLTU)}, 2014, pp. 32--38.

\bibitem{adel2015syntactic}
H.~Adel, N.~T. Vu, K.~Kirchhoff, D.~Telaar, and T.~Schultz, ``Syntactic and
  semantic features for code-switching factored language models,''
  \emph{IEEE/ACM transactions on audio, speech, and language Processing},
  vol.~23, no.~3, pp. 431--440, 2015.

\bibitem{adel2013combination}
H.~Adel, N.~T. Vu, and T.~Schultz, ``Combination of recurrent neural networks
  and factored language models for code-switching language modeling,'' in
  \emph{Proc. Annual Meeting of the Association for Computational (ACL)}, 2013,
  pp. 206--211.

\bibitem{vu2012first}
N.~T. Vu, D.-C. Lyu, J.~Weiner, D.~Telaar, T.~Schlippe, F.~Blaicher, E.-S.
  Chng, T.~Schultz, and H.~Li, ``A first speech recognition system for
  mandarin-english code-switch conversational speech,'' in \emph{Proc.
  International Conference on Acoustics, Speech and Signal Processing
  (ICASSP)}.\hskip 1em plus 0.5em minus 0.4em\relax IEEE, 2012, pp. 4889--4892.

\bibitem{yilmaz2018acoustic}
E.~Y{\i}lmaz, H.~v.~d. Heuvel, and D.~A. van Leeuwen, ``Acoustic and textual
  data augmentation for improved asr of code-switching speech,'' in \emph{Proc.
  Annual Conference of the International Speech Communication Association
  (INTERSPEECH)}.\hskip 1em plus 0.5em minus 0.4em\relax {ISCA}, 2018, pp.
  1933--1937.

\bibitem{guo2018study}
P.~Guo, H.~Xu, L.~Xie, and E.~S. Chng, ``Study of semi-supervised approaches to
  improving english-mandarin code-switching speech recognition,'' in
  \emph{Proc. Annual Conference of the International Speech Communication
  Association (INTERSPEECH)}, 2018, pp. 1928--1932.

\bibitem{zeng2018end}
Z.~Zeng, Y.~Khassanov, V.~T. Pham, H.~Xu, E.~S. Chng, and H.~Li, ``On the
  end-to-end solution to mandarin-english code-switching speech recognition,''
  in \emph{Proc. Annual Conference of the International Speech Communication
  Association (INTERSPEECH)}, 2019, pp. 2165--2169.

\bibitem{luo2018towards}
N.~Luo, D.~Jiang, S.~Zhao, C.~Gong, W.~Zou, and X.~Li, ``Towards end-to-end
  code-switching speech recognition,'' in \emph{Proc. Annual Conference of the
  International Speech Communication Association (INTERSPEECH)}, 2020, pp.
  4776--4780.

\bibitem{tian2022lae}
J.~Tian, J.~Yu, C.~Zhang, C.~Weng, Y.~Zou, and D.~Yu, ``Lae: Language-aware
  encoder for monolingual and multilingual asr,'' in \emph{Proc. Annual
  Conference of the International Speech Communication Association
  (INTERSPEECH)}, 2022, pp. 3178--3182.

\bibitem{song2022monolingual}
T.~Song, Q.~Xu, H.~Lu, L.~Wang, H.~Shi, Y.~Lin, Y.~Yang, and J.~Dang,
  ``Monolingual recognizers fusion for code-switching speech recognition,''
  \emph{arXiv preprint arXiv:2211.01046}, 2022.

\bibitem{song2022language}
T.~Song, Q.~Xu, M.~Ge, L.~Wang, H.~Shi, Y.~Lv, Y.~Lin, and J.~Dang,
  ``Language-specific characteristic assistance for code-switching speech
  recognition,'' in \emph{Proc. Annual Conference of the International Speech
  Communication Association (INTERSPEECH)}, 2022, pp. 3924--3928.

\bibitem{dalmia2021transformer}
S.~Dalmia, Y.~Liu, S.~Ronanki, and K.~Kirchhoff, ``Transformer-transducers for
  code-switched speech recognition,'' in \emph{Proc. International Conference
  on Acoustics, Speech and Signal Processing (ICASSP)}.\hskip 1em plus 0.5em
  minus 0.4em\relax IEEE, 2021, pp. 5859--5863.

\bibitem{yan2022towards}
B.~Yan, M.~Wiesner, O.~Klejch, P.~Jyothi, and S.~Watanabe, ``Towards zero-shot
  code-switched speech recognition,'' \emph{arXiv preprint arXiv:2211.01458},
  2022.

\bibitem{zhang2021rnn}
S.~Zhang, J.~Yi, Z.~Tian, J.~Tao, and Y.~Bai, ``Rnn-transducer with language
  bias for end-to-end mandarin-english code-switching speech recognition,'' in
  \emph{Proc. International Symposium on Chinese Spoken Language Processing
  (ISCSLP)}.\hskip 1em plus 0.5em minus 0.4em\relax IEEE, 2021, pp. 1--5.

\bibitem{li2019towards}
K.~Li, J.~Li, G.~Ye, R.~Zhao, and Y.~Gong, ``Towards code-switching asr for
  end-to-end ctc models,'' in \emph{Proc. International Conference on
  Acoustics, Speech and Signal Processing (ICASSP)}.\hskip 1em plus 0.5em minus
  0.4em\relax IEEE, 2019, pp. 6076--6080.

\bibitem{zhang2022reducing}
S.~Zhang, J.~Yi, Z.~Tian, J.~Tao, Y.~T. Yeung, and L.~Deng, ``Reducing language
  context confusion for end-to-end code-switching automatic speech
  recognition,'' \emph{arXiv preprint arXiv:2201.12155}, 2022.

\bibitem{peng2022minimum}
Y.~Peng, J.~Zhang, H.~Xu, H.~Huang, and E.~S. Chng, ``Minimum word error
  training for non-autoregressive transformer-based code-switching asr,'' in
  \emph{Proc. International Conference on Acoustics, Speech and Signal
  Processing (ICASSP)}.\hskip 1em plus 0.5em minus 0.4em\relax IEEE, 2022, pp.
  7807--7811.

\bibitem{chuang2021non}
S.-P. Chuang, H.-J. Chang, S.-F. Huang, and H.-y. Lee, ``Non-autoregressive
  mandarin-english code-switching speech recognition,'' in \emph{Proc.
  Automatic Speech Recognition and Understanding Workshop (ASRU)}.\hskip 1em
  plus 0.5em minus 0.4em\relax IEEE, 2021, pp. 465--472.

\bibitem{peng2022internal}
Y.~Peng, Y.~Liu, J.~Zhang, H.~Xu, Y.~He, H.~Huang, and E.~S. Chng, ``Internal
  language model estimation based language model fusion for cross-domain
  code-switching speech recognition,'' in \emph{Proc. Annual Conference of the
  International Speech Communication Association (INTERSPEECH)}, 2022, pp.
  1666--1670.

\bibitem{lee2020modeling}
G.~Lee and H.~Li, ``Modeling code-switch languages using bilingual parallel
  corpus,'' in \emph{Proc. Annual Meeting of the Association for Computational
  Linguistics (ACL)}, 2020, pp. 860--870.

\bibitem{zhou2020multi}
X.~Zhou, E.~Y{\i}lmaz, Y.~Long, Y.~Li, and H.~Li, ``Multi-encoder-decoder
  transformer for code-switching speech recognition,'' in \emph{Proc. Annual
  Conference of the International Speech Communication Association
  (INTERSPEECH)}, 2020, pp. 1042--1046.

\bibitem{yang2002acoustic}
Y.~Yang and B.~Wang, ``Acoustic correlates of hierarchical prosodic boundary in
  mandarin,'' in \emph{Proc. Speech prosody international conference}, 2002.

\bibitem{dong2020cif}
L.~Dong and B.~Xu, ``Cif: Continuous integrate-and-fire for end-to-end speech
  recognition,'' in \emph{Proc. International Conference on Acoustics, Speech
  and Signal Processing (ICASSP)}.\hskip 1em plus 0.5em minus 0.4em\relax IEEE,
  2020, pp. 6079--6083.

\bibitem{sennrich2015neural}
R.~Sennrich, B.~Haddow, and A.~Birch, ``Neural machine translation of rare
  words with subword units,'' in \emph{Proc. Annual Meeting of the Association
  for Computational (ACL)}, 2016, pp. 1715--1725.

\bibitem{gulati2020conformer}
A.~Gulati, J.~Qin, C.-C. Chiu, N.~Parmar, Y.~Zhang, J.~Yu, W.~Han, S.~Wang,
  Z.~Zhang, Y.~Wu \emph{et~al.}, ``Conformer: Convolution-augmented transformer
  for speech recognition,'' in \emph{Proc. Annual Conference of the
  International Speech Communication Association (INTERSPEECH)}, 2020, pp.
  5036--5040.

\bibitem{srivastava2014dropout}
N.~Srivastava, G.~Hinton, A.~Krizhevsky, I.~Sutskever, and R.~Salakhutdinov,
  ``Dropout: a simple way to prevent neural networks from overfitting,''
  \emph{The journal of machine learning research}, vol.~15, no.~1, pp.
  1929--1958, 2014.

\bibitem{lyu2010seame}
D.-C. Lyu, T.-P. Tan, E.~S. Chng, and H.~Li, ``Seame: a mandarin-english
  code-switching speech corpus in south-east asia,'' in \emph{Proc. Annual
  Conference of the International Speech Communication Association
  (INTERSPEECH)}, 2010, pp. 1986--1989.

\bibitem{vaswani2017attention}
A.~Vaswani, N.~Shazeer, N.~Parmar, J.~Uszkoreit, L.~Jones, A.~N. Gomez,
  {\L}.~Kaiser, and I.~Polosukhin, ``Attention is all you need,'' in
  \emph{Proc. Advances in Neural Information Processing Systems (NIPS)}, 2017,
  pp. 5998--6008.

\bibitem{kingma2014adam}
D.~P. Kingma and J.~Ba, ``Adam: A method for stochastic optimization,'' in
  \emph{Proc. International Conference on Learning Representations (ICLR)},
  2015.

\bibitem{graves2012sequence}
A.~Graves, ``Sequence transduction with recurrent neural networks,'' in
  \emph{Proc. International Conference on Machine Learning (ICML)}, 2012.

\end{thebibliography}

\end{document}